# Reaction-Drift Model for Switching Transients in $Pr_{0.7}Ca_{0.3}MnO_3$-Based Resistive RAM

Vivek Saraswat[1], Shankar Prasad[1], Abhishek Khanna, Ashwin Wagh, Ashwin Bhat, Neeraj Panwar, Sandip Lashkare and Udayan Ganguly

*Department of Electrical Engineering, Indian Institute of Technology Bombay, Mumbai – 400076, India*

*Abstract*— $Pr_{0.7}Ca_{0.3}MnO_3$ (PCMO) based RRAM shows promising memory properties like non-volatility, low variability, multiple resistance states and scalability. From a modeling perspective, the charge carrier DC current modeling of PCMO RRAM by drift diffusion (DD) in the presence of fixed oxygen ion vacancy traps and self-heating (SH) in Technology Computer Aided Design (TCAD) (but without oxygen ionic transport) was able to explain the experimentally observed space charge limited conduction (SCLC) characteristics, prior to resistive switching. Further, transient analysis using DD+SH model was able to reproduce the experimentally observed fast current increase at ~100 ns timescale, prior to resistive switching. However, a complete quantitative transient current transport *plus* resistive switching model requires the inclusion of ionic transport. We propose a Reaction-Drift (RD) model for oxygen ion vacancy related trap density variation, which is combined with the DD+SH model. Earlier we have shown that the Set transient consists of 3 stages and Reset transient consists of 4 stages experimentally. In this work, the DD+SH+RD model is able to reproduce the entire transient behavior over 10 ns - 1 s range in timescale for both the Set and Reset operations for different applied biases and ambient temperatures. Remarkably, a universal Reset experimental behavior, $log(I) \propto (m \times log(t))$ where $m \approx -1/10$ is reproduced in simulations. This model is the first model for PCMO RRAMs to significantly reproduce transient Set/Reset behavior. This model establishes the presence of self-heating and ionic-drift limited resistive switching as primary physical phenomena in these RRAMs.

## I. INTRODUCTION

$Pr_{0.7}Ca_{0.3}MnO_3$ (PCMO) is used in a non-filamentary resistive switching memory device (RRAM). It is attractive due to better variability and multi-level resistance states [1], [2]. A forming-less operation is observed in PCMO [3], [4], which simplifies the memory controller. The device is a W/PCMO/Pt i.e. a metal-oxide-metal structure fabricated on an $SiO_2/Si$ substrate. The PCMO film (60 nm) is deposited using a room temperature pulsed laser deposition followed by annealing at 650°C. The top contact (W) size is ~ 1 μm. The detailed fabrication methods are explained earlier in [5]. From a mechanisms' perspective, in PCMO based RRAM, the following extent of understanding exists in literature. To explain the experimental DC IV characteristics, Space Charge Limited Current (SCLC) mechanism has been invoked for current transport [6]–[8]. Resistance is modulated by trap-density – consistent with trap SCLC [9]. We have presented a simple trap density extraction methodology based on trap SCLC model to correlate trap density change with resistance switching [10]. Further, a Technology Computer Aided Design (TCAD) model consisting of drift-diffusion (DD) based holes transport in p-type semiconductor with self-heating (SH) (but without ionic transport) to model SCLC current is able to replicate DC IV characteristics at lower bias, i.e. prior to onset of resistive switching for a range of ambient temperatures (25°C - 125°C) [11]. The inclusion of self-heating enabled the replication of non-linear behavior, earlier erroneously attributed to Trap-Filled Limit [9]. The signature of self-heating was further confirmed by fast (sub-100ns) transient switching behavior [5]. Transient TCAD modeling was able to match the experimental current transient prior to the onset of resistance switching.

The switching phenomena in filamentary RRAMs like $HfO_x$ based RRAMs and the corresponding quantitative modelling of the steady-state and transient measurements has received widespread attention from researchers [12], [13], [22]–[24], [14]–[21]. The switching mechanisms in uniform switching (bulk conduction) RRAMs has also been explored extensively [25], [26], [35]–[41], [27]–[34]. The effects of top electrode properties [30] and an interface layer formation during switching [39] have also been demonstrated. Although, there is a large amount of experimental data available for switching in these films, a quantitative model for the current transients in uniform switching RRAMs involving ion dynamics is still lacking. A qualitative explanation of resistive switching in PCMO based RRAM is as follows. Resistive switching in PCMO based RRAM is related to the transport of oxygen ion (or equivalently oxygen vacancies) [42]–[45]. Reversible ionic transport occurs by reversing bias polarity to drift ions to and from a reactive electrode (i.e. an oxygen source/sink) via lattice substitutions to modulate oxygen vacancy concentration in PCMO [46]–[48]. These oxygen vacancies are related to hole traps (defects) in the lattice [10]. Thus, ionic transport modulates bulk trap concentration to produce resistance modulation of current under trap SCLC mechanism [9].

[1]Vivek Saraswat and Shankar Prasad have equal contributions.
Revision submitted on 15-Apr-20.
Correspondence e-mail: udayan@ee.iitb.ac.in

This work was supported by Centre of Excellence in Nanofabrication – IITB, Prime Minister's Research Fellowship, Ministry of Electronics and Information Technology.

While such a *qualitative* model has been presented earlier, the detailed dynamics of Set/Reset transients needs to be explored and *quantitatively* modeled. Recently, we have experimentally studied the short to long range (10 ns – 1 s) transient to highlight the signature of ion dynamics for Set/Reset (Fig.1). All pulse measurements were performed on the Agilent B1530A waveform generator/fast measurement unit (WGFMU) [5]. A three-stage Set is observed due to a step bias input - initial current increase (S1) is followed by current saturation (S2), which is followed by abrupt current increase to compliance (S3). A four-stage Reset is observed where an initial increase in current (R1) is followed by a fast decrease in current (R2) to a current saturation level (R3).

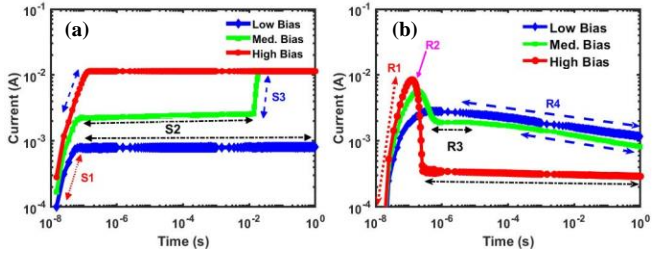

Fig.1. Experimental (a) Set (b) RESET transients are shown at various applied bias. For Set, a three-stage transient is observed from a step voltage input - (S1) initial current increase is followed by (S2) current saturation, followed by (S3) abrupt current increase to compliance. For RESET, a four-stage transient is observed from a step voltage input - (R1) an initial current increase in current is followed by (R2) a fast current decrease to (R3) a current saturation level. Finally, (R4) a slow universal current reduction of $I \propto t^{-\frac{1}{10}}$, over 6 orders of magnitude in time.

Finally, a slow universal current reduction (R4) of $I \propto t^{-\frac{1}{10}}$ is observed over 6 orders of magnitude in time. Such a specific and multi-stage behavior is attractive for quantitative model development to demonstrate detailed understanding of the switching mechanism.

In this paper, we introduce an ionic Reaction-Drift (RD) model to include ion dynamics, coupled with our prior TCAD based DD+SH model [11]. We show that the model quantitatively replicates the experimental Set and Reset Write current transients for a range of biases and at different ambient temperature (300-450K). Particularly interesting are the reproduction of slow universal Reset behavior and different Set vs. Reset voltage-time dilemma behaviors. Thus, such an analysis enables a quantitative understanding of resistive switching mechanism and the primary physical phenomena operating in PCMO based RRAM.

## II. Transient Resistive Switching Model

Earlier, the DD+SH model (without ionic transport/reaction) was able to capture the short timescale transient response (stage S1), which was dependent upon self-heating timescale, to saturation (stage S2) [5]. However ionic transport occurs for Set where current increases sharply (stage S3). The deviation of DD+SH simulations from experiment occurs in stage S3 as the model lacks the ionic transport and vacancy generation model and hence, it cannot account for resistive switching. Experimentally for Set stage S3, ionic motion leads to a current increase, which increases Joule heating in the device (i.e. self-heating) to further increase ionic motion. Thus, a positive feedback mechanism is set up to create a sudden sharp increase in current to compliance (Fig. 1(a)). The ionic transport, indicated by the ionic drift velocity ($v_{drift}$) at a given temperature and electric field, is given by Mott-Gurney Equation [49]:

$$v_{drift} = a.f.\exp\left(-\frac{E_m}{k_B T}\right).\sinh\left(\frac{\xi}{\xi_0}\right) \quad (1)$$

where $a$ is the hopping distance, $f$ is the escape frequency, $\xi$ is the electric field, $\xi_0 = 2k_B T/qa$ is the characteristic electric field and $E_m$ is the activation barrier, $k_B$ is the Boltzmann constant, $T$ is the absolute device temperature. Ionic motion timescale is essentially the timescale at which this $v_{drift}$ produces significant ionic motion compared to present trap density. As the current is measured in log scale from 10 ns to 1 s, the positive feedback is observed as a sudden sharp take off in current towards compliance when *the measurement timescale matches the ionic motion timescale* for a given applied bias (stage S3). As bias is reduced, we observe an exponentially longer timescale in saturation (stage S2) to abrupt current take-off (stage S3) i.e. occurrence of ion motion. This is qualitatively consistent with (1). Essentially, lower field and lower current cause lower heating and consequently lower temperature. Lower electric field ($\xi$) and temperature ($T$), will reduce $v_{drift}$ to increase timescale for equivalent ionic transport distance exponentially as shown in (1). To validate this qualitative explanation, a quantitative model based on numerical simulations is presented in this paper.

The Reset transient phenomenon has 3 different applied voltage regimes (Fig. 1(b) – low, med, high). The high bias has a fast increase (stage R1) and then fast decrease in current (stage R2) followed by a time-independent current (stage R3). For intermediate bias, the high bias behavior is observed, but the constant current (stage R3) is higher. Eventually, *when the measurement timescale matches the ionic motion timescale,* the constant current starts to reduce to follow the "universal" curve with the time exponent (log(I)-log(t) plot) of approximately -1/10 (stage R4). The low bias shows initial current increase (stage R1) followed by the "universal" current transient curve with the time exponent of -1/10 (stage R4). We first model this specific "universal" behavior. To begin, we show that a universal power law dependence is possible in an isothermal case where ionic transport increases trap density and hence reduces current – similar to Negative Bias Temperature Instability (NBTI) analysis in MOSFETs [50], [51]. However, this universal power law derived using isothermal assumption has an inaccurate (i.e. larger than observed "-1/10") time exponent. Inclusion of self-heating in the device rectifies this error as discussed later.

For Reset, an increase in hole trap density ($N_T$) reduces current, $I_{trap,SCLC}$ [10], [11] as given by:

$$I_{trap,SCLC} \sim \frac{I_{trap,free}}{\left(\frac{N_T}{N_V}\right)\exp\frac{E_T - E_V}{k_B T}} \propto \frac{1}{N_T} \quad (2)$$

where $I_{trap,free}$ is the trap free SCLC current, $N_T$ is the hole trap density, $N_V$ is the hole effective density of states, $E_T$ is the trap energy level and $E_V$ is the valence band edge. We assume two concurrent processes: (a) the applied bias attracts the anions throughout the PCMO bulk towards the reactive electrode (W) and the ionic/vacancy motion occurs in a substitutional manner (Fig. 2), (b) the anions at the W-PCMO interface observe a sink and undergo redox activity to create a source of new vacancies/traps (reaction in Fig. 2).

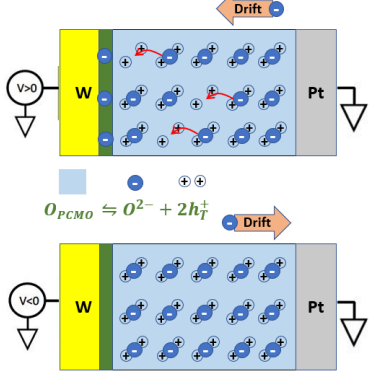

Fig. 2. Schematic view of generation and movements of ions and vacancies inside the PCMO device in presence of applied electric field.

To create positive traps at the interface, a substitutional unit element ($A_{lattice}$) produces an anion ($A^{n-}$) with $n$ charges and a vacancy that yields $n$ hole traps ($h_T^+$) per anion (3). This anion sinks at the reactive electrode and any excess electrons escape through the contact (4):

$$A_{lattice} \leftrightharpoons A^{n-} + nh_T^+ \quad (3)$$
$$W + xA^{n-} \leftrightharpoons WA_x + x.n\,e^- \quad (4)$$

The bias polarity-based drift is essential to explain bipolarity in bipolar RRAM. Under applied electric field, $[A^{n-}]$ can drift towards the reactive electrode as governed by (1). We assume first order chemical kinetics for each reactant in (3). We also assume that the entire process is drift limited, i.e. the reaction (3) is essentially close to equilibrium, which means the forward and backward reaction rates are approximately equal:

$$n\frac{d[h_T]}{dt} = \frac{d[A^{n-}]}{dt} = k_F[A_{lattice}] - k_R[A^{n-}][h_T]^n \approx 0 \quad (5)$$
$$\text{constant} = k_F[A_{lattice}] \approx k_R[A^{n-}][h_T]^n \quad (6)$$
$$[A^{n-}] \approx \frac{k_{eq}}{[h_T]^n} \quad (7)$$

where $k_{eq} = k_F[A_{lattice}]/k_R$. This is a constant because $[A_{lattice}]$ is almost constant for small extent of reaction, and $k_F$ and $k_R$ are also constants under isothermal condition. Also, these $A^{n-}$ ions at the reactive electrode are instantaneously consumed (4). The total ionic concentration leaving PCMO is $[A^{n-}]$ concentration of ions drifting a small distance $dl = v_{drift} \times dt$. For simplicity, we assume a uniform change in concentration, given by,

$$d[A^{n-}] = \frac{[A^{n-}]v_{drift}dt}{L} \quad (8)$$
$$\frac{d[A^{n-}]}{dt} = \frac{v_{drift}[A^{n-}]}{L} = n\frac{d[h_T]}{dt} \quad (9)$$

where $L$ = PCMO film thickness. Using (7), we substitute out $[A^{n-}]$ in (9) to construct a differential equation in the hole trap density, $[h_T]$,

$$\frac{d[h_T]}{dt} = \frac{k_{eq}v_{drift}}{nL[h_T]^n} \quad (10)$$

Integrating, we get the time evolution of hole trap density:

$$[h_T] = \left(\frac{n+1}{n} \times \frac{k_{eq}v_{drift}}{L}\right)^{\frac{1}{n+1}} t^{\frac{1}{n+1}} \quad (11)$$

Assuming $[h_T] = N_T$ i.e uniform trap density, then (1) gives:

$$I_{trap,SCLC} \propto t^{-\frac{1}{n+1}} \quad (12)$$

Thus, $n$ traps per anion $A^{n-}$, based on Reaction-Drift (RD) model, would produce a time exponent $m = -1/(n+1)$. The experimental exponent of $\sim -1/10$ requires $n = 9$. Oxygen is the only anion in PCMO. It is indeed quite difficult that a single diffusing species of oxygen produces 9 traps. The first possibility is an oxygen ion ($O^{2-}$), which produces $n = 2$ traps is widely reported which should produce an exponent of $m = -1/3$, much sharper than the observed $-1/10$. Next, a superoxide ion ($O_2^-$) should produce $n = 4$ traps. However, superoxide ions are unstable and only reported for surface diffusion and dissociation into $O^{2-}$ ions for bulk diffusion in solid oxide fuel cell (SOFC) electrode studies of LSMO [52]. Thus, for an isothermal case, it is highly unlikely to get $n = 9$.

Next, we show qualitatively that self-heating should reduce the magnitude of this time exponent. For Reset, the device starts in low resistance state. Upon application of bias, the current increases with self-heating (stage R1). Consequently, the temperature rises, and ion transport is initiated to increase trap density, which reduces the current (stage R2). First, in the case of high bias, a very high current is reached to enable a high temperature quickly. At high temperature, rapid ion transport occurs and saturates at a high value to reduce the current abruptly resulting in abrupt temperature drop too. Trap density and consequently current become time independent. Second, in the case of lower bias, a lower final current is reached (stage R1). Consequently, the temperature rises to a moderate level (lower than the high bias case), where ionic transport increases trap density slowly to reduce the current (stage R4). The gentle current reduction will reduce the temperature gently, which will further reduce the ionic transport. A negative feedback occurs due to temperature reduction, which will slow down ionic transport in time compared to the isothermal case. Naturally, the rate of current reduction will also slow down. Thus, the magnitude of power law exponent will reduce from the isothermal prediction of 1/3. We will show that we can achieve the exponent of $m = -1/10$ using detailed DD+SH+RD numerical simulations. Based on this model, we also show that high bias regime i.e. fast increase and then fast decrease of current followed by constant current is also realized. Further, the intermediate bias regime i.e. initially high bias like behavior (stage R1 and R2), which results in a higher constant current will remain unchanged in time (stage R3) as the initial timescale of measurement is too fast for ionic transport at that intrinsic device temperature. However, at longer timescale of measurement when the timescale becomes comparable to the rate of ionic transport at that device temperature, then the transient current reduces again and merges with the "universal" curve of time exponent $m = -1/10$ (stage R4).

## III. SIMULATION SETUP

In this section, we describe the time dependent hole current transients simulation methodology for a fixed applied bias, ambient temperature and starting trap density (low resistance state, low $N_T$ - LRS or high resistance state, high $N_T$ - HRS) (Fig. 3). After the initialization in device temperature, $T$, trap density, $N_T$ and voltage ramp time, the detailed time evolution is performed in MATLAB at the fixed applied bias. For updating the trap density during simulation transient, the Reaction-Drift model based on Mott Gurney ionic drift (1) and the ionic reaction kinetics (10) explained in the Section II are used. In addition, the device temperature is calculated as a function of the input power density and the thermal parameters. The hole current is obtained from a look-up-table (LUT) of quasi-static simulations performed in Sentaurus TCAD for different fixed trap density values and isothermal device temperature values. The following sections explain the simulation steps in detail:

### A. Assumptions

Before describing the time dynamics, it is essential to understand three assumptions involved:
a) Trap density is uniform across device (no spatial $N_T$ dependence),
b) Temperature is uniform across device (lumped or point-device model assumption) and,
c) Hole current response time is much faster than temperature or ionic Reaction-Drift response time.

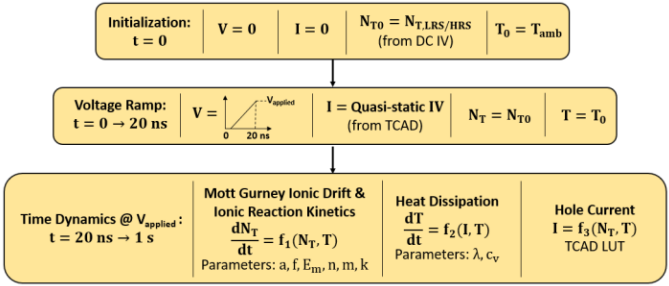

Fig. 3. Flowchart describing update of electric current I, trap density $N_T$, and device temperature T during initialization, ramp and fixed voltage time dynamics solved in MATLAB. Function $f_1$ refers to (10), $f_2$ refers to (13) – Section III.D.2 and $f_3$ is shown in Fig. 4.

It is true that ionic transport may lead to non-uniform trap density. Still, assumption (a) is reasonable because of two observations. Firstly, we have earlier shown by simulations that asymmetry in trap spatial profile only weakly affects the largely symmetric IV characteristics [10]. Secondly, the experimental IV characteristics are essentially symmetric for positive vs. negative bias before the onset of ionic transport i.e. Set/Reset [10]. Thus, a uniform trap density is a reasonable approximation. For the assumption (b), detailed finite element method (FEM) thermal simulations were performed for the device stack earlier [5], [11] to extract equivalent thermal capacitance and conductivity values for a lumped or point device model to be used in this work. Hence, assumption (b) is a useful simplification which allows capturing the temperature dependence in resistive switching without the simulation complexity of a more complete spatial Fourier heat equation. We have also ignored the capacitive response delays of the hole currents (assumption (c)), because firstly, holes have high mobility compared to ions due to significantly smaller mass. Secondly, the objective is to capture the long timescale transients governed by temperature and ionic Reaction-Drift phenomena and the hole response time is expected to be much shorter.

### B. Initialization

The total time simulation has four variables: (1) external applied bias, $V(t)$, this is the input; (2) carrier current, $I(t)$; (3) trap density, $N_T(t)$ and (4) device temperature, $T(t)$. At t = 0, the start of the transient simulation, the current and voltage are zero and the device is in one of LRS ($N_{T0} \sim 10^{18}$ /$cm^3$) or HRS ($N_{T0} \sim 10^{20}$ /$cm^3$) state. It is possible to extract the starting trap density, $N_{T0}$ from a given resistance state as observed in experimental DC IV measurements [10]. The device is in equilibrium with the ambient temperature to begin with ($T = T_{amb}$).

### C. Voltage Ramp

The input bias rises from 0 to $V_{app}$ in a short span of 20 ns (same as experiment). This timescale is much shorter to result in any significant change in trap density or device temperature. Hence, for this duration, these values are held at their initialized values and the hole current is updated in time from TCAD quasi-static simulation values corresponding to the voltage sweep values (0 to $V_{app}$).

### D. Time Dynamics at applied bias

The time evolution that follows happens at a fixed external bias ($V_{app}$). In this duration, the current, temperature and trap density evolution is inter-dependent. In Section II, the time dynamics of the trap density (10) was discussed using a Reaction-Drift model. The drift velocity of the oxygen ions follows the Mott-Gurney law and this velocity is utilized to evaluate the ionic reaction kinetics. As a whole, the Reaction-Drift model takes the present trap density and device temperature as inputs. The device temperature is a simple heat balance between input power and dissipation at any moment in time. Since the hole response time is small compared to temperature and traps transients, a quasi-static LUT generated in TCAD for different $T$ and $N_T$ can be used immediately to calculate hole current at any instant in time.

*1) Current Transport in TCAD:*

The current LUT is generated using quasi-static IV simulations in Sentaurus TCAD (Fig. 4). In current transport, TCAD uses electron-hole drift-diffusion formalism to compute current through the PCMO film ($L$ = 65 nm) with the experimental top contact area (1 $um^2$) at a fixed uniform trap density ($N_T$) and device ambient temperature ($T_{amb}$). The device is isothermally in equilibrium with ambient temperature ($T = T_{amb}$) and trap density does not evolve during TCAD current calculation. This is repeated for several $T_{amb}$ and $N_T$ values. Essentially the TCAD model solves the Poisson, carrier continuity and carrier statistics self-consistently. The current values are stored for a fixed value of applied voltage as a function of $N_T$ and $T$. This constitutes an LUT for that applied bias to be used later in the MATLAB solver. We had earlier demonstrated excellent matching of temperature dependent current transport in PCMO based RRAM using trap SCLC model with self-heating [10] implemented in Sentaurus [22]. Further, the dc model was extended to perform transient simulations where transient current for fast Set and Reset was modeled to show excellent match with experimental current transients before the on-set of ionic transport [11]. Our TCAD solver and material parameter files are hence calibrated to experimentally observed current conduction except the resistive switching phenomena for which we propose the Reaction-Drift model external to TCAD LUT generation in this work.

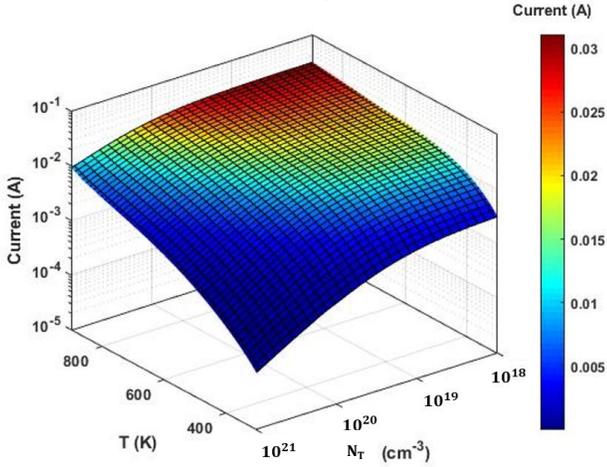

Fig. 4. Look-up Table (LUT) for $I(N_T, T)$ for $V_{app}$=1.5 V, $T_{amb}$ = 300 K generated using isothermal and fixed trap density quasistatic IV simulations performed in TCAD for different device temperatures and trap densities.

*2) Thermal Model:*

For a single point device, the transient device temperature, $T$ can be described in terms of its thermal properties as:

$$\frac{(T - T_{amb})}{R_{th}} + c_s \frac{dT}{dt} = I.V \quad (13)$$

where $R_{th}$ = thermal resistance of device ($R_{th} = \frac{L}{A\lambda}$, $L$ = device conduction length, $A$ is the device area and $\lambda$ = effective thermal conductivity of device), $T_{amb}$ = ambient temperature, $c_s$ = heat capacitance ($c_s = c_v AL$, $c_v$ is the effective specific heat capacity of device). This device temperature affects the current conduction through the device by means of the LUT and affects the trap density based on the Reaction-Drift model discussed in Section II.

## IV. RESULTS AND DISCUSSIONS

The self-consistent solver based on models and simulation setup described in Sections II and III is used to reproduce the experimental characteristics significantly (parameters used are listed in Table I) as discussed in the following sections.

TABLE I
PARAMETERS USED IN THE SOLVER

| Model | Symbol | Quantity | Value |
|---|---|---|---|
| Reaction Kinetics | $k_{eq}$ | $k_F[A_{lattice}]/k_R$ | $\sim(2 /nm^3)^3$ |
| | $n$ | Traps per anion | 2 |
| Mott Gurney Ionic Drift | $a$ | Hopping distance | 0.5 nm |
| | $f$ | Escape frequency | $1 \times 10^{13}$ Hz |
| | $E_m$ | Migration barrier | 0.8 eV |
| Thermal Model | $c_v$ | Specific heat capacity | $2 \times 10^7$ J/m³K |
| | $\lambda$ | Thermal Conductivity | ~6 W/mK |
| Device | $L$ | PCMO film thickness | 65 nm |
| | $A$ | Top contact area | 1 μm × 1 μm |
| Inputs | $V_{app}$ | Applied bias | $\pm[1.3 - 2.5]$ V |
| | $T_{amb}$ | Ambient temperature | $300 - 475$ K |

### A. Effect of n on Reset Transient

The Reset transient for $n = 1, 2$ and 4 is shown in Fig. 5. Essentially, without self-heating, we obtain a $m = -1/(n + 1)$ time exponent. However, with self-heating, the magnitude of the time exponent is reduced. The Reaction-Drift model along with self-heating can thus be instrumental in predicting the transient Reset timescales for materials with different mobile ionic species.

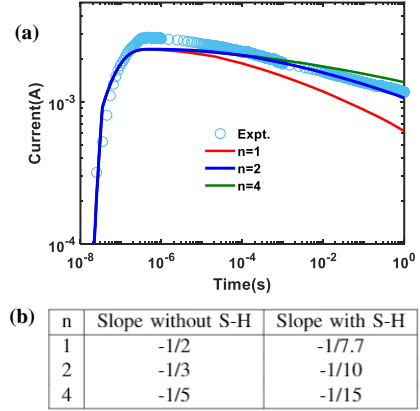

Fig. 5. (a) Reset current vs time shows slow switching with different slopes for n=1, 2 and 4 (b) Different slope values during slow switching for n=1, 2 and 4 with and without self-heating (SH)

### B. Reset Simulations

The effect of bias on Reset simulations is shown in Fig. 6. At high bias, $T(t)$ increases quickly (~ 800 K). $N_T$ rises fast and saturates (Fig. 6 (a)). Accordingly, the three regimes are observed i.e. (i) current increases quickly and (ii) then decreases quickly to (iii) saturation state (Fig. 6(b)). At low bias, the temperature rises to ~ 400 K, the associated change in trap density is small too (Fig. 6(c)). Then a slow negative feedback between temperature and trap density occurs. Accordingly, two regimes in current transient is observed: (i) current increases and then (ii) follows the universal time exponent (Fig. 6(d)).

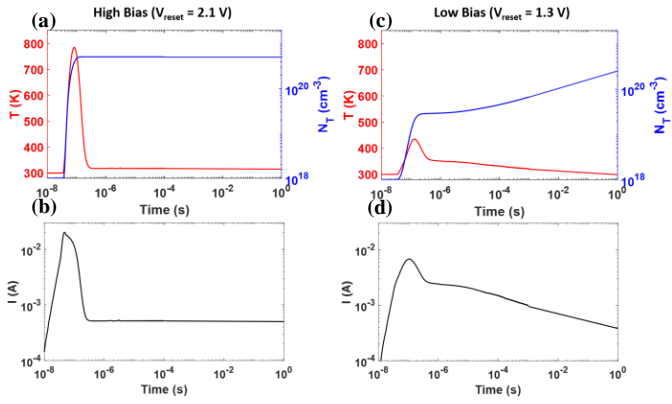

Fig. 6. Reset Transients. Simulated (a) Temperature and Trap density vs time and (b) Current vs time shows fast switching for high bias. Simulated (c) Temperature and Trap density vs time and (d) Current vs time and shows slow switching for low bias

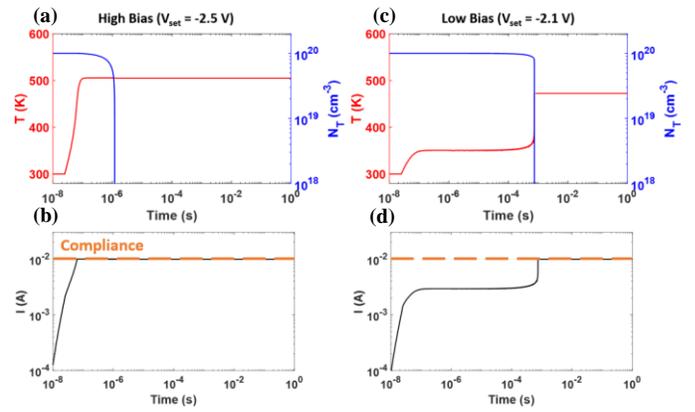

Fig. 8. Set Transients. Simulated (a) Temperature and Trap density vs time and (b) Current vs time shows fast switching for high bias. Simulated (c) Temperature and Trap density vs time and (d) Current vs time and shows slow switching for low bias

Fig. 7 (a) compares simulated vs. experimental $I(t)$ for a range of biases. Our model is able to reproduce the experimental behavior quite comprehensively. Small quantitative differences can be attributed to the various simplifying assumptions (Section III. A). Next, we simulate the effect of ambient temperature ($T_{amb}$). Our transient simulations are in excellent agreement for a range of $T_{amb}$ (i.e. 300-450K). We observe that at high-bias (yellow-red curves in Fig. 7(a), (b)), there is a saturation to indicate that higher bias does not produce faster switching beyond 100ns timescale. Further, at low bias, the switching timescale is also limited 10 ms. $T_{amb}$ does not strongly affect these levels. Simulations show excellent agreement with experiments (Fig. 7 (c), (d) and (e)).

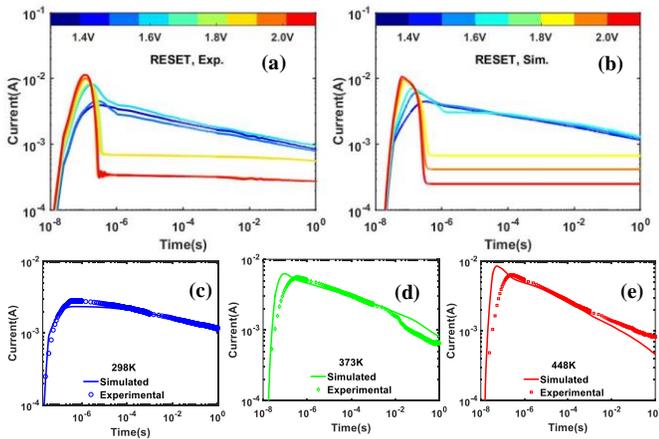

Fig. 7. Reset, (a) Experiment (b) Simulation shows qualitative matching of Current vs Time during Reset for range of step voltage input. (c), (d), (e) Matching of slow switching of current for different $T_{amb}$ (300-450K) at $V_{app} = 1.5$ V.

C. *Set Simulations*

The effect of bias on Set simulations is shown in Fig. 8. At high bias, the temperature (Fig. 8(a)) and current (Fig. 8(b)) rise sharply to compliance. $N_T(t)$ reduces quickly too (Fig. 8(a)). For low bias, temperature (Fig. 8(c)) and the current (Fig. 8(d)) increases then saturates. Then $N_T(t)$ falls at a specific timescale when the current and accordingly the temperature rise sharply (Fig. 8(c), (d)) to compliance. Set simulations include a compliance current level (~ 10 mA, as in measurements). Beyond compliance, the experiment *shows* compliance current as fixed (which is true) but with the desired voltage applied (which is not actual). The actual voltage is reduced appropriately to ensure current compliance. However, the actual voltage is not displayed in output by the instrumentation. In simulations as well, the output current is actual (compliance) but the voltage is still the set point not the reduced actual value. As Set time is defined as time to reach a critical current level below compliance, the experiment and simulations for $T$ or $N_T$ beyond compliance is not of interest in the present study.

Next, we show that the experimental Set transient is captured well by simulations at 300K as seen in Fig. 9 (a), (b). A constant current (10 mA) for simulations and experiment is used to extract Set time as a function of applied bias for $T = 300$ K, 375 K, 450 K in Fig. 9 (c). At higher bias, Set time shows saturation and is essentially is limited to 100ns. At lower bias, there is a fast increase in Set time, which is essentially the voltage-time dilemma. This is in excellent agreement with simulations.

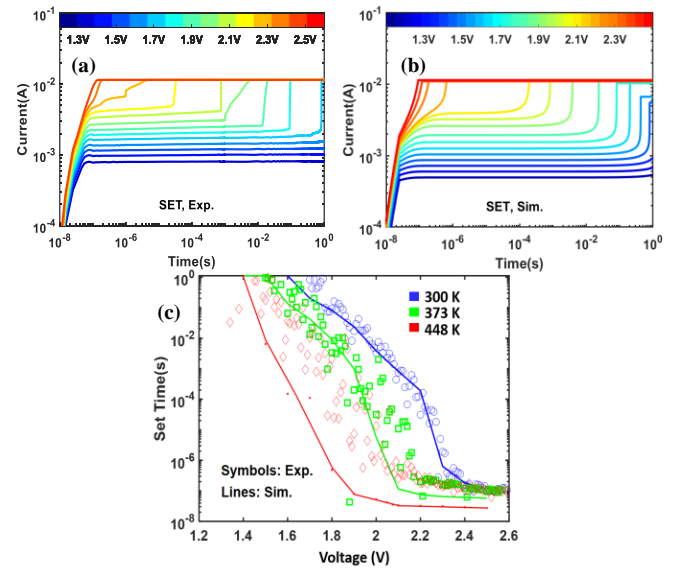

Fig. 9. Set, (a) Experiment (b) Simulation shows qualitatively matching of Current vs time during Set for different step voltage input. (c) Set time vs Voltage for three different $T_{amb}$ (i.e. 300-450K)

Further, the $T_{amb}$ dependence on the Set behavior shows that the voltage-time dilemma at low bias responds to temperature. However, at high bias, the timescale of Set is still limited to 100ns, which is temperature insensitive. These simulation results are also in excellent agreement with experiments (Fig. 9(c)). This effect of ambient temperature on improving Write timescales at low biases is more pronounced for Set (Fig. 9(c)) as compared to Reset (Fig. 7(c), (d) and (e)). We expect thermal design of RRAM to become an important consideration. Thermally insulating design for DC performance enhancement has been experimental demonstrated [53] but still remains to be seen for the transient performance. Further, PCMO composition engineering has been performed to show tradeoff between speed and retention [54]. Reaction-Drift model captures the device temperature in comparison to the ambient temperature using effective thermal conductance and capacitance of the RRAM stack. We expect the results of switching times at different ambient temperatures in this work (Fig. 9(c)) to translate to thermally efficient RRAM stacks operating at room temperature. We expect the predictions from this model to enable optimal composition design in addition to thermal design.

## V. Conclusion

In this paper, a Reaction-Drift Model is proposed to include ion dynamics to a drift-diffusion with self-heating-based model for hole transport. We demonstrate that the model can reproduce experimentally observed Set and Reset transient across a range of timescale (100 ns – 1 s), Set/Reset bias and ambient temperatures (300 – 450 K). Remarkably, a universal Reset behavior of time-exponent $m \approx -1/10$ is replicated. The simulations are able to capture the difference between Set/Reset timescales vs Set/Reset voltages to explain the different voltage-time dilemma observed in Set vis a vis Reset. The ambient temperature shows a stronger effect for Set compared to Reset – which is captured well in simulations. Further, the timescale for fast switching is limited to ~ 100 ns for Set and Reset – which is independent of ambient temperature is also captured. Thus, we present a simple point device model of Set/Reset in PCMO based RRAM that can comprehensively reproduce timescale, bias and temperature effects. Such models with further refinements will enable a detailed understanding of physical phenomena and design of PCMO based RRAMs.